\documentclass[prc,twocolumn,showpacs]{revtex4}
\usepackage{bm}
\usepackage{amssymb}
\usepackage{amsmath}
\usepackage{graphicx}
\begin{document}
%%%%%%%%%%%
%%%%%%%%%%%
\title{Weak charge form factor and radius of $^{208}$Pb through parity violation in electron scattering}
%%%

\affiliation{Carnegie Mellon University, Pittsburgh, Pennsylvania 15213, USA} %CMU  
\affiliation{INFN, Sezione di Roma, I-00161 Rome, Italy} %INFN Roma
\affiliation{University of Tennessee, Knoxville, TN, and Indiana University, Bloomington, Indiana 47405, USA} %IUB
\affiliation{Syracuse University, Syracuse, New York 13244, USA} %SU
\affiliation{Thomas Jefferson National Accelerator Facility, Newport News, Virginia 23606, USA}  %JLab
\affiliation{University of Massachusetts Amherst, Amherst, Massachusetts  01003, USA} %UMass
\affiliation{University of Virginia, Charlottesville, Virginia  22903, USA} %UVa
\author{C.~J.~Horowitz}\email{horowit@indiana.edu}\affiliation{University of Tennessee, Knoxville, TN, and Indiana University, Bloomington, Indiana 47405, USA} %IUB
\author{Z.~Ahmed}\affiliation{Syracuse University, Syracuse, New York 13244, USA} % SU
\author{M.~M.~Dalton}\affiliation{University of Virginia, Charlottesville, Virginia  22903, USA} \author{G.~B.~Franklin}\affiliation{Carnegie Mellon University, Pittsburgh, Pennsylvania  15213, USA} %CMU
\author{M.~Friend}\affiliation{Carnegie Mellon University, Pittsburgh, Pennsylvania  15213, USA} %CMU
\author{C.-M.~Jen}\affiliation{Syracuse University, Syracuse, New York 13244, USA} % SU
\author{K.~S.~Kumar}\affiliation{University of Massachusetts Amherst, Amherst, Massachusetts  01003, USA} %UMass
\author{N.~Liyanage}\affiliation{University of Virginia, Charlottesville, Virginia  22903, USA} %UVa
\author{D.~McNulty}\altaffiliation[now at ]{Idaho State University, Pocatello, Idaho  83209, USA}\affiliation{University of Massachusetts Amherst, Amherst, Massachusetts  01003, USA}
\author{J.M.~Mammei}\altaffiliation[now at ]{University of Manitoba, Winnipeg, Canada}\affiliation{University of Massachusetts Amherst, Amherst, Massachusetts  01003, USA} %UMass
\author{L.~Mercado}\affiliation{University of Massachusetts Amherst, Amherst, Massachusetts  01003, USA} %UMass
\author{R.~W.~Michaels}\affiliation{Thomas Jefferson National Accelerator Facility, Newport News, Virginia  23606, USA} %JLab
\author{K.~D.~Paschke}\affiliation{University of Virginia, Charlottesville, Virginia  22903, USA} %UVa
\author{A.~Rakhman}\affiliation{Syracuse University, Syracuse, New York 13244, USA} % SU
\author{S.~Riordan}\altaffiliation[previously at ]{University of Virginia, Charlottesville, Virginia  22903, USA}
\affiliation{University of Massachusetts Amherst, Amherst, Massachusetts  01003, USA} %UMass
\author{B. Quinn}\affiliation{Carnegie Mellon University, Pittsburgh, Pennsylvania 15213, USA} %CMU  
\author{K.~Saenboonruang}\affiliation{University of Virginia, Charlottesville, Virginia  22903, USA} %UVa
\author{R.~Silwal}\affiliation{University of Virginia, Charlottesville, Virginia  22903, USA} %UVa
\author{P.~A.~Souder}\affiliation{Syracuse University, Syracuse, New York 13244, USA} % SU
\author{G.~M.~Urciuoli}\affiliation{INFN, Sezione di Roma, I-00161 Rome, Italy} %INFN Roma}
\author{J.~Wexler}\affiliation{University of Massachusetts Amherst, Amherst, Massachusetts  01003, USA} %UMass

%\author{C. J. Horowitz}\email{horowit@indiana.edu} 
%\affiliation{Physics Division, Oak Ridge National Laboratory, P.O. Box 2008, Oak Ridge, TN 37831, USA, \\
%and Center for Exploration of Energy and Matter and Department of Physics,
%             Indiana University, Bloomington, IN 47405, USA}
%\author{R. Michaels}
%\affiliation{Thomas Jefferson National Accelerator Facility Newport News, VA, USA}

%%%%%%%%%%%
\date{\today}
\begin{abstract}
We use distorted wave electron scattering calculations to extract the weak charge form factor $F_W(\bar q)$, the weak charge radius $R_W$, and the point neutron radius $R_n$, of $^{208}$Pb from the PREX parity violating asymmetry measurement.  The form factor is the Fourier transform of the weak charge density at the average momentum transfer $\bar q=0.475$ fm$^{-1}$.  We find $F_W(\bar q) =0.204 \pm 0.028 ({\rm exp}) \pm 0.001 ({\rm model})$.  We use the Helm model to infer the weak radius from $F_W(\bar q)$.  We find $R_W= 5.826 \pm 0.181 ({\rm exp}) \pm 0.027 ({\rm model})\ {\rm fm}$.  Here the exp error includes PREX statistical and systematic errors, while the model error describes the uncertainty in $R_W$ from uncertainties in the surface thickness $\sigma$ of the weak charge density.    The weak radius is larger than the charge radius, implying a ``weak charge skin'' where the surface region is relatively enriched in weak charges compared to (electromagnetic) charges.  We extract the point neutron radius $R_n=5.751 \pm 0.175\ ({\rm exp}) \pm 0.026 ({\rm model}) \pm 0.005 ({\rm strange})\ {\rm fm}$, from $R_W$.  Here there is only a very small error (strange) from possible strange quark contributions.  We find $R_n$ to be slightly smaller than $R_W$ because of the nucleon's size.    Finally, we find a neutron skin thickness of $R_n-R_p=0.302\pm 0.175\ ({\rm exp}) \pm 0.026$ (model) $\pm$ 0.005 (strange) fm, where $R_p$ is the point proton radius.
\end{abstract}
\smallskip
\maketitle

%\section{Introduction}
%Nuclear charge densities have been accurately measured with elastic electron scattering and have become our picture of the atomic nucleus, see for example ref. \cite{chargeden}.  These measurements have had an enormous impact.  Unfortunately, neutron densities are not directly probed in electron scattering because the neutron is uncharged.

Parity violating elastic electron scattering provides a model independent probe of neutron densities, because the weak charge of a neutron is much larger than the weak charge of a proton \cite{dds}.  In Born approximation, the parity violating asymmetry $A_{pv}$, the fractional difference in cross sections for positive and negative helicity electrons, is proportional to the weak form factor $F_W$.  This is very close to the Fourier transform of the neutron density.  Therefore the neutron density can be extracted from an electro-weak measurement \cite{dds}.  However, one must include the effects of Coulomb distortions, which have been accurately calculated \cite{couldist}, if the charge density $\rho_{ch}$ \cite{chargeden} is well known.  Many details of a practical parity violating experiment to measure neutron densities, along with a number of theoretical corrections, were discussed in a long paper \cite{bigpaper}.
%However, the Born approximation is not valid for a heavy nucleus and coulomb distortion effects must be included.  These were calculated in ref. \cite{couldist} by numerically solving the Dirac equation for an electron scattering in both the coulomb potential and a weak interaction axial vector potential.  Many details of a practical parity violating experiment to measure neutron densities have been discussed in a long paper \cite{bigprex}.    

Recently, the Lead Radius Experiment (PREX) measured $A_{pv}$ for 1.06 GeV electrons, scattered by about five degrees from $^{208}$Pb, and the neutron radius $R_n$ was extracted \cite{PREX1}.        To do this, the experimental $A_{pv}$ was compared to a least squares fit of $R_n$ as a function of $A_{pv}$, predicted by seven mean field models \cite{sban} (see also \cite{roca}).  In the present paper, we provide a second, more detailed, analysis of the measured $A_{pv}$.  This second analysis provides additional information, such as the weak form factor, and clarifies the (modest) model assumptions necessary to extract $R_n$. 

We start with distorted wave calculations of $A_{pv}$ for an electron moving in Coulomb and weak potentials \cite{couldist}.  We use these to extract the weak form factor from the PREX measurement.  In Born approximation, one can determine the weak form factor directly from the measured $A_{pv}$.  However, Coulomb distortions may make $A_{pv}$ sensitive to the weak form factor for a range of momentum transfers $q$.  In addition, the experimental acceptance for PREX includes a range of momentum transfers for laboratory scattering angles from about 3.5 to 8 degrees \cite{PREX1}.  Therefore we will need to make very modest assumptions about the shape of the weak form factor (how it depends on momentum transfer $q$) in order to determine the value of the form factor at the average momentum transfer $\bar q$\ \cite{PREX1},
\begin{equation}
 \bar q= \langle Q^2 \rangle^{1/2} = 0.475 \pm 0.003\ {\rm fm}^{-1}.
 \label{q2}
 \end{equation}

We initally assume the weak charge density of $^{208}$Pb, $\rho_W(r)$ has a Wood Saxon form,
\begin{equation}
\rho_W(r) = \frac{\rho_0}{1 + {\rm exp}[(r-R)/a]},
\label{rhoW}
\end{equation}
with parameters $\rho_0$, $R$ and $a$.  Note, this form is only used to access the sensitivity to the shape of the form factor and our results will be independent of this assumed form.  The weak density is normalized to the weak charge $Q_W=\int d^3r \rho_W(r)$, see below.  

We define the weak form factor $F_W(q)$ as the Fourier transform of $\rho_W(r)$, 
\begin{equation}
F_W(q)=\frac{1}{Q_W}\int d^3r \frac{\sin q r}{q r} \rho_W(r).
\label{F(q)}
\end{equation}
This is normalized $F_W(q=0)=1$.  Our procedure is to calculate $A_{pv}(\theta)$, including full Coulomb distortions \cite{couldist}, assuming $\rho_W$ from Eq. \ref{rhoW}.  We average $A_{pv}(\theta)$ over laboratory scattering angle $\theta$ using the experimental acceptance $\epsilon(\theta)$\ \cite{PREX1},
\begin{equation}
\langle A \rangle = \frac{\int d\theta \sin\theta\, \epsilon(\theta) \frac{d\sigma}{d\Omega} A_{pv}} {\int d\theta \sin\theta\, \epsilon(\theta) \frac{d\sigma}{d\Omega}}.
\label{<A>}
\end{equation}
Here the unpolarized elastic cross section is $\frac{d\sigma}{d\Omega}$.  We then adjust $R$ until the calculated $\langle A\rangle$ agrees with the PREX result \cite{PREX1}  
\begin{equation}
A^{Pb}_{pv}=0.656 \pm 0.060 ({\rm stat}) \pm 0.014 ({\rm syst}) \ {\rm ppm}.
\label{Apv} 
\end{equation}
Here the first error is statistical and the second error includes systematic contributions.  For $a=0.6$ fm, we obtain a central value of $R=6.982$ fm, see below.  Finally from the $\rho_W(r)$ in Eq. \ref{rhoW}, that reproduces $A^{Pb}_{pv}$, we calculate $F_W(\bar q)$ using Eq. \ref{F(q)}.  This procedure fully includes Coulomb distortions and depends slightly on the assumed surface thickness $a$ in Eq. \ref{rhoW}.   In Table \ref{Table1} we show Wood Saxon fits to seven nonrelativistic and relativistic mean field model weak charge densities considered in ref. \cite{sban}.  Note that these models span a very large range of neutron radii $R_n$.  The average value of $a$ for these models is $0.61\pm 0.05$ fm.  Using a central value of $a=0.6$ fm we obtain,
\begin{equation}
F_W(\bar q) =0.204 \pm 0.028 ({\rm exp}) \pm 0.001 ({\rm mod}).
\label{F}
\end{equation}
Here the first experimental error is from adding the statistical and systematic errors in Eq. \ref{Apv} in quadrature.  The second model error is from varying  $a$ by $\pm 0.05$ fm.  This shows that the extracted form factor is all but independent of the assumed shape of the weak charge density.  Equation \ref{F} is a major result of this paper.  This is the form factor of the weak charge density that is implied by the PREX measurement.

\begin{table}
\caption{Least squares fits of Wood Saxon ($R$, $a$, see Eq. \ref{rhoW}) or Helm model ($R_0$, $\sigma$, see Eq. \ref{helmrho}) parameters to theoretical mean field model weak charge densities.} 
\begin{tabular}{lllll}
 &\multicolumn{2}{c}{ Wood Saxon} & \multicolumn{2}{c}{Helm} \\
Mean field force & $R$ (fm) & $a$ (fm)& $R_0$ (fm)& $\sigma$ (fm) \\
\toprule
Skyrme I \cite{SI} & 6.655 & 0.564 & 6.792 & 0.943 \\
Skyrme III \cite{SIII} & 6.820 & 0.613 & 6.976 & 1.024 \\
Skyrme SLY4 \cite{SLY4} & 6.700 & 0.668  & 6.888 & 1.115 \\
FSUGold \cite{FSUgold} & 6.800 & 0.618 & 6.961 & 1.028 \\
NL3 \cite{NL3} & 6.896 & 0.623 & 7.057 & 1.039 \\
NL3p06 \cite{sban}& 6.730 & 0.606  & 6.886 & 1.010 \\
NL3m05 \cite{sban} &7.082 & 0.605  & 7.231 & 1.012 \\
\toprule
Average & & 0.61 $\pm$ 0.05 & &1.02$\pm$ 0.09 \\
\end{tabular} 
\label{Table1}
\end{table}

We now explore some of the implications of Eq. \ref{F} using the Helm model \cite{HELM}  for the weak form factor.  In the past, the Helm model has proven very useful for analyzing (unpolarized) electron scattering form factors \cite{HELM2,HELM3}, see also ref. \cite{HELM4} for an application of the Helm model to neutron rich nuclei.  The weak charge density is first assumed to be uniform out to a diffraction radius $R_0$.  This uniform density is then folded with a gaussian of width $\sigma$ to get the final weak density.  The width $\sigma$ includes contributions from both the surface thickness of the point nucleon densities and the single nucleon form factor.  In the Helm model, the weak form factor has a very simple form,
\begin{equation}
F_W(q)=\frac{3}{q R_0} j_1(qR_0) {\rm e}^{-\sigma^2q^2/2},
\label{helmf}
\end{equation} 
with $j_1(x)=\sin x/x^2-\cos x/x$ a spherical Bessel function.  The diffraction radius $R_0$ determines the location $q_0$ of the zero in the weak form factor $F_W(q_0)=0$.  In coordinate space, the Helm model weak charge density can be written in terms of error functions (erf),
\begin{align}
\rho_W(r)= {}& \frac{3Q_W}{8\pi R_0^3} \Bigl\{{\rm erf}\bigl(\frac{R_0+r}{\sqrt{2}\sigma}\bigr)-{\rm erf}\bigl(\frac{r-R_0}{\sqrt{2}\sigma}\bigr) \nonumber\\ 
{}& + \sqrt{\frac{2}{\pi}} \frac{\sigma}{r}\bigl({\rm e}^{-\frac{1}{2}(\frac{r+R_0}{\sigma})^2} - {\rm e}^{-\frac{1}{2}(\frac{r-R_0}{\sigma})^2} \bigr) \Bigr\}.
\label{helmrho}
\end{align}
The root mean square radius of the weak charge density $R_W$ (or weak radius) is
$R_W^2=\int d^3 r\, r^2\, \rho_W(r)/Q_W$,
\begin{equation}
R_W^2=\frac{3}{5}\bigl(R_0^2 + 5 \sigma^2 \bigr).
\label{RW}
\end{equation}
We see that Eq. \ref{F} implies via Eq. \ref{helmf} a relationship between allowed values of $R_0$ and $\sigma$.  This relationship then implies via Eq. \ref{RW} a range of weak radii.  Thus Eq. \ref{F} does not, by itself, determine the weak radius.  In principle the rms radius follows from the derivative of the form factor with respect to $Q^2$ at $q=0$.  Because the PREX measurement is at finite $q$, one needs to assume some information about the surface thickness $\sigma$ in order to extract $R_W$.  Alternatively within the Helm model, if one determined the location of the zero of the form factor $q_0$, in addition to Eq. \ref{F}, then this would uniquely fix both $R_0$ and $\sigma$ and so determine $R_W$. 

In Table \ref{Table1} we collect values of $\sigma$ determined by least squares fits of the Helm density, Eq. \ref{helmrho}, to seven model mean field densities.  The average of $\sigma$ for the seven mean field densities is 1.02 fm and individual results deviate by no more than 0.09 fm from this average.  If one assumes $\sigma=1.02\pm 0.09$ fm, Eqs. \ref{F}, \ref{helmf} and \ref{RW} imply
\begin{equation}
R_W=5.826 \pm 0.181 ({\rm exp}) \pm 0.027 ({\rm mod})\ {\rm fm}.
\label{RWresult}
\end{equation} 
Again the larger experimental (exp) error is from adding the statistical and systematic errors in Eq. \ref{Apv} in quadrature, while the model (mod) error comes from the {\it coherent} sum of the assumed $\pm$ 0.09 fm uncertainty in $\sigma$ and the $\pm 0.001$ model error in $F_W$.  The model error in Eq. \ref{RWresult} provides an estimate of the uncertainty in $R_W$ that arrises because of uncertainties in the surface thickness.  Of course, it is not guaranteed that all theoretical models will have a surface thickness within the range $1.02\pm 0.09$ fm.  Nevertheless, this result suggests that uncertainties in surface thickness are much less important for $R_W$ than either the present PREX experimental error or even that of an improved measurement where the experimental error is reduced by about a factor of three \cite{PREXII}.   This is consistent with earlier results of Furnstahl \cite{furnstahl} suggesting a nearly unique relation between $F_W(\bar q)$ and the point neutron radius $R_n$.  We emphasize that if uncertainties in the surface thickness are a concern, one should compare theoretical predictions for the form factor $F_W(\bar q)$ to Eq. \ref{F}, instead of comparing theoretical predictions for $R_W$ to Eq. \ref{RWresult}.

Comparing Eq. \ref{RWresult} to the experimental charge radius $R_{ch}=5.503$ fm \cite{chargeden,chargeradius} implies a ``weak charge skin'' of thickness
\begin{equation}
R_W-R_{ch}=0.323 \pm 0.181 ({\rm exp}) \pm 0.027 ({\rm mod}) \ {\rm fm}.
\label{weakskin}
\end{equation}
Thus the surface region of $^{208}$Pb is relatively enhanced in weak charges compared to electromagnetic charges.   This weak charge skin is closely related to the expected neutron skin, see below.  Equation \ref{weakskin}, itself, represents an experimental milestone.  We now have direct evidence that the weak charge density, of a heavy nucleus, is more extended than the electromagnetic charge density.   

\begin{figure}[ht]
\begin{center}
\includegraphics[width=3.7in,angle=0,clip=true] {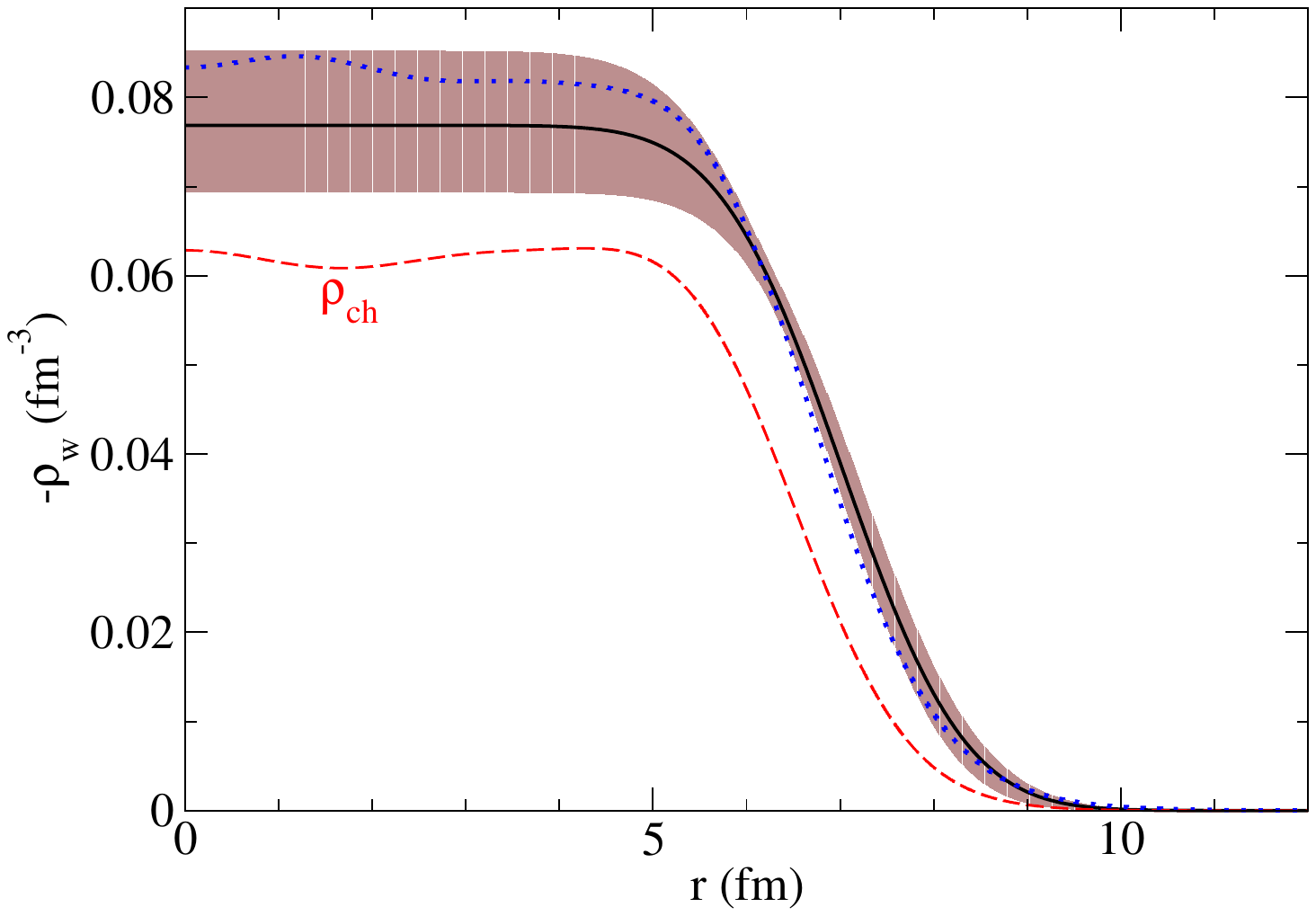}
\caption{(Color on line) Helm model weak charge density $-\rho_W(r)$ of $^{208}$Pb that is consistent with the PREX result (solid black line).  The brown error band shows the incoherent sum of experimental and model errors.   The red dashed curve is the experimental (electromagnetic) charge density $\rho_{ch}$ and the blue dotted curve shows a  sample mean field result based on the FSUGold interaction \cite{FSUgold}.}
\label{Fig1}
\end{center}
\end{figure}

In Fig. \ref{Fig1} we show a Helm model weak charge density that is consistent with the PREX measurement.  This figure shows an uncertainty range from the experimental error and a model uncertainty from the assumed $\pm 0.09$ fm uncertainty in $\sigma$.   Parameters for these densities are presented in Table \ref{Table2}.  We also show in Fig. \ref{Fig1} the (electromagnetic) charge density \cite{chargeden} and a typical mean field weak charge density based on the FSUGold interaction, see Eq. \ref{rhoWfinal} below.  This theoretical density is within the error bars of the Helm model density.     

\begin{table}
\caption{Helm model weak charge density parameters $R_0$ and $\sigma$ that reproduce the following values for the weak form factor $F_W(\bar q)$, see Eqs. \ref{F} and \ref{helmf}.} 
\begin{tabular}{llll}
Density & $R_0$ (fm) & $\sigma$ (fm)& $F_W(\bar q)$\\
\toprule
Central value & 7.167 & 1.02 & 0.204\\
Exp error bar & 7.417 & 1.02 & 0.176\\
Exp error bar & 6.926 & 1.02 & 0.232\\
Model error bar & 7.137 & 1.11& 0.203\\
Model error bar & 7.194 & 0.93 & 0.205\\
\end{tabular} 
\label{Table2}
\end{table}

Finally we wish to extract $R_n$ for $^{208}$Pb from $R_W$ in Eq. \ref{RWresult}.  We start by reviewing the relationship between the point proton radius $R_p$ and the measured charge radius $R_{ch}$.  Ong et al. have \cite{Ong}
\begin{equation}
R_{ch}^2=R_p^2 + \langle r_p^2 \rangle + \frac{N}{Z}\langle r_n^2\rangle + \frac{3}{4M^2}+ \langle r^2 \rangle_{so}.
\end{equation}
Here the charge radius of a single proton is $\langle r_p^2\rangle=0.769$ fm$^2$ and that of a neutron is $\langle r_n^2\rangle=-0.116$ fm$^2$.  We calculate that the contribution of spin-orbit currents to $R_{ch}$ is small because of cancelations between protons and neutrons $\langle r^2\rangle_{so}=-0.028$ fm$^2$.  Finally the Darwin contribution $3/4M^2$ is also small with $M$ the nucleon mass.  For $^{208}$Pb we have,
$
R_{ch}^2=R_p^2+0.5956\ {\rm fm}^2,
$
or, for $R_{ch}=5.503$ fm \cite{chargeden,chargeradius},
\begin{equation}
R_p=5.449\ {\rm fm}.
\end{equation}

For the weak charge density of a spin zero nucleus, we neglect meson exchange and spin-orbit currents and write \cite{bigpaper}
\begin{equation}
\rho_W(r)=4\int d^3r'\bigl[G_n^Z(|{\bf r}-{\bf r'}|)\rho_n(r')+G_p^Z(|{\bf r}-{\bf r'}|)\rho_p(r')\bigr]\, .
\label{rhoWinitial}
\end{equation}
Here the density of weak charge in a single proton $G_p^Z(r)$ or neutron $G_n^Z(r)$ is the Fourier transform of the nucleon (Electric) Sachs form factors $G_p^Z(Q^2)$ and $G_n^Z(Q^2)$.  These describe the coupling of a $Z^0$ boson to a proton or neutron \cite{bigpaper},
\begin{equation}
4G_p^Z=q_pG_E^p+q_n G_E^n -G_E^s,
\end{equation}
\begin{equation}
4G_n^Z=q_nG_E^p + q_p G_E^n -G_E^s.
\end{equation}
At tree level, the weak nucleon charges are $q^0_n=-1$ and $q_p^0=1-4\sin^2\Theta_W$.  We include radiative corrections by using the values $q_n=-0.9878$ and $q_p=0.0721$ based on the up $C_{1u}$ and down $C_{1d}$ quark weak charges in refs.  \cite{qweak,pdb}. The Fourier transform of the proton (neutron) electric form factor is $G_E^p(r)$ ($G_E^n(r)$) and has total charge  $\int d^3r G_E^p(r)=1$  ($\int d^3r G_E^n(r)=0$).  Finally $G_E^s$ describes strange quark contributions to the nucleon's electric form factor \cite{SAMPLEsff,HAPPEXsff,G0sff,A4sff}.  Note that there may be some small uncertainty regarding the $Q^2$ dependence of the radiative corrections.  This uncertainty could change $R_n^2$, see below, by a very small amount of order $(1+q_n)\langle r_p^2\rangle$.

Equation \ref{rhoWinitial} can be rewritten by using a similar expression for $\rho_{ch}$
\begin{equation}
\rho_W(r)=q_p\, \rho_{ch}(r) + \int d^3r'\bigl[q_n ( G_E^p\rho_n + G_E^n\rho_p) - G_E^s\rho_b\bigr]\, 
\label{rhoWfinal}
\end{equation}
with $\rho_b=\rho_n+\rho_p$.  The weak charge of $^{208}$Pb is
\begin{equation}
Q_W=\int d^3r \rho_W(r)=Nq_n+Zq_p=-118.55.
\label{QWEAK}
\end{equation}
From Eq. \ref{rhoWfinal}, we relate the point neutron rms radius $R_n$, to $R_W$,
\begin{equation}
R_n^2=\frac{Q_W}{q_nN}R_W^2 -\frac{q_p Z}{q_n N} R_{ch}^2 -\langle r_p^2 \rangle -\frac{Z}{N}\langle r_n^2\rangle +\frac{Z+N}{q_nN}\langle r_s^2\rangle,
\end{equation}
where $\langle r_s^2\rangle=\int d^3r' r'^2 G_E^s(r')$ is the square of the nucleon strangeness radius.  
This yields
\begin{equation}
R_n^2 = 0.9525 R_W^2 - 1.671 \langle r_s^2\rangle + 0.7450\ {\rm fm}^2.
\end{equation}
The strangeness radius of the nucleon $\langle r_s^2\rangle^{1/2}$  is constrained by experimental data \cite{SAMPLEsff,HAPPEXsff,G0sff,A4sff} and their global analysis \cite{YoungSff,LiuSff}.  Using Table V of ref. \cite{LiuSff} for $Q^2<0.11$ GeV$^2$, gives $\langle r_s^2\rangle = -6 dG_E^s/dQ^2=0.02 \pm 0.04 \approx \pm 0.04$ fm$^2$.

The neutron radius then follows from Eq. \ref{RWresult},
\begin{equation}
R_n=5.751 \pm 0.175\ ({\rm exp}) \pm 0.026 ({\rm mod})\pm 0.005 ({\rm str})\ {\rm fm}\, .
\label{Rn}
\end{equation}
Here the very small third (str) error is from possible strange quark contributions.  The neutron radius $R_n$ is slightly smaller than $R_W$ because of the nucleon's size.  Finally, the neutron skin thickness is
\begin{equation}
R_n-R_p=0.302\pm 0.175({\rm exp}) \pm 0.026 ({\rm mod}) \pm 0.005 ({\rm str})\ {\rm fm}.
\end{equation}
This result agrees, within the model error, with the result of ref. \cite{PREX1}, $R_n-R_p=0.33^{+0.16}_{-0.18}$\ fm.  The small difference between the present result and ref. \cite{PREX1} arrises because of small limitations of the Helm model in representing theoretical mean field densities.  For example the Helm model does not have the correct expodential behavior at large distances.  However, we have clarified how the extraction of the neutron radius depends upon assumptions on the weak skin thickness $\sigma$ and we provide an explicit model error for the uncertainty in $R_n-R_p$ because of uncertainties in $\sigma$.

We now summarize our results.  In this paper we use distorted wave electron scattering calculations for $^{208}$Pb to extract the weak charge form factor $F_W(\bar q)$, Eq. \ref{F}, the weak radius $R_W$, Eq. \ref{RWresult}, and the point neutron radius $R_n$, Eq. \ref{Rn}, from the PREX parity violating asymmetry measurement.  The weak form factor is the Fourier transform of the weak charge density at the average momentum transfer of the experiment.  This quantity is essentially model independent and is insensitive to assumptions about the surface thickness.

The extraction of $R_W$ depends on modest assumptions about the surface thickness.  We use the Helm model to derive an estimate on the uncertainty in $R_W$ because of the uncertainty in surface thickness.  We find a ``weak charge skin'' where the surface region is relatively enriched in weak charges compared to electromagnetic charges.  This is closely related to the neutron skin where $R_n$ is larger than the point proton radius $R_p$.  Finally, we extract $R_n$, given $R_W$, and find it to be slightly smaller than $R_W$ because of the nucleon's size.

We thank Witek Nazarewicz for very helpful discussions.  We gratefully acknowledge the hospitality of the University of Tennessee and the Physics Division of ORNL where this work was started.  This work was supported in part by DOE grant DE-FG02-87ER40365.

%%%%%%%%%%%%%%%%%%%%%%%%%%%%%%%%%%%%%%%%%%%%%%%%%%%%%%%%%%%%%%%%%
\vfill\eject

\end{document}